\documentclass[reprint,amsmath,amssymb,aps,]{revtex4-2}

\usepackage{subfig}
\usepackage{tikz}
\usepackage{quantikz}
\usepackage{graphicx}
\usepackage{dcolumn}
\usepackage{bm}
\usepackage{hyperref}
\usepackage{algorithm}
\usepackage{algorithmic}
\usepackage{mathtools}
\usepackage{todonotes}
\DeclarePairedDelimiter{\ceil}{\lceil}{\rceil}

\newenvironment{mydefinition}[1]{\textbf{#1}:}{}

\newenvironment{myexample}[1]{\textbf{Example #1}:}{}
\begin{document}

\preprint{APS/123-QED}

\title{Quantum Relaxation for Solving Multiple Knapsack Problems}

\author{Monit SHARMA$^{1}$, Jin YAN$^{1}$}
\author{Hoong Chuin LAU$^{1,2}$}
\email{Corresponding author email: hclau@smu.edu.sg}
\address{$^1$School of Computing and Information Systems,  Singapore Management University, Singapore}
\address{$^2$Institute of High Performance Computing, A*STAR, Singapore}

\author{Rudy RAYMOND$^{3}$}
\email{Current affiliation is JP Morgan Chase \& Co., USA}
\address{$^3$IBM Quantum, IBM Japan}

\begin{abstract}
Combinatorial problems are a common challenge in business, requiring finding optimal solutions under specified constraints. While significant progress has been made with variational approaches such as QAOA, most problems addressed are unconstrained (such as Max-Cut). In this study, we investigate a hybrid quantum-classical method for constrained optimization problems, particularly those with knapsack constraints that occur frequently in financial and supply chain applications. Our proposed method relies firstly on relaxations to local quantum Hamiltonians, defined through commutative maps. Drawing inspiration from quantum random access code (QRAC) concepts, particularly Quantum Random Access Optimizer (QRAO), we explore QRAO's potential in solving large constrained optimization problems. We employ classical techniques like Linear Relaxation as a presolve mechanism to handle constraints and cope further with scalability. We compare our approach with QAOA and present the final results for a real-world procurement optimization problem: a significant-sized multi-knapsack-constrained problem.

\end{abstract}

\maketitle

\section{\label{sec:level1}Introduction}

The concept of leveraging quantum computers to generate approximate solutions for NP-hard combinatorial problems dates back more than two decades, initially introduced through quantum adiabatic eigenstate evolution \cite{QAEE}. This foundational idea evolved into the Quantum Approximate Optimization Algorithm (QAOA) \cite{QAOA1} and Variational Quantum Eigensolver (VQE)\cite{VQE}, which entails variational optimization of quantum parameters \cite{QAOA2}. With considerable interest in harnessing quantum advantage for classical combinatorial problems, extensive research has scrutinized the performance of QAOA \cite{warmstart, Zhou_2020, tate2022bridging}. These quantum optimization techniques rely on establishing a bijective mapping between the space of classical binary variables and logical basis states of a collection of qubits, as originally outlined \cite{QAOA1}. Within the QAOA framework, the cost function optimized on a quantum computer typically features a classical maximal eigenstate, which doesn't strictly necessitate superposition and entanglement for preparation. This contrasts with quantum many-body Hamiltonians, where extremal eigenstates often exhibit significant entanglement. In such cases, quantum computers benefit from a natural memory advantage in storing the ground state.

QAOA and VQE, as classical-quantum hybrid algorithms for near-term quantum devices, face critical scalability issues due to their encoding of one classical bit per qubit. The limited qubit capacity of current quantum devices significantly restricts the size of problem instances that can be addressed, hindering the applicability of these algorithms to larger-scale computational challenges.

QRAO \cite{QRAO} offers a solution to this challenge by encoding multiple classical bits (typically three or fewer) into one qubit. By doing so, it facilitates the generation of approximate solutions for combinatorial problems that seek extremal eigenstates of local quantum Hamiltonians. These local quantum Hamiltonians represent relaxations of the original combinatorial problems. For each element in the image of a combinatorial cost function, it becomes feasible to construct a quantum state with the same Hamiltonian expectation value. This integration between classical combinatorial optimization and quantum relaxation is a key feature of QRAO, making it a promising approach for addressing complex optimization challenges using quantum-inspired techniques~\cite{Kondoh2024}. Recent results of QRAO give hints of its robustness to quantum noise~\cite{Tamura2024}, and its power to leverage quantum entanglement to obtain optimal and better solutions~\cite{Teramoto2023}. However, the problems addressed by QRAO were all unconstrained optimization. 

In this study, our objective is to address a complex constrained supply chain problem using QRAO. We will explore the effectiveness of QRAO in addressing the inherent complexities of constrained supply chain problems, by solving a Multiple Knapsack Problem (MKP) and comparing it with the well-studied QAOA approach which is effective in solving unconstrained problems such as Max-Cut \cite{zhou2020quantum}.
We will scale a real-world multiple knapsack-based Risk-Aware Procurement Optimization problem involving over 100 variables, demonstrating the potential of combining QRAO with Linear Relaxation (LR) \cite{LR}. This work merges quantum and classical computation to enhance the utility of quantum algorithms, demonstrating the scalability of QRAO and its efficacy in handling larger problem sizes, where traditional methods like QAOA are limited by qubit and memory requirements.

The subsequent sections of this paper are structured
as follows:
In Section. \ref{sec:level2} we described the formulation for MKP and the Risk-Aware Procurement Optimization problem, followed by Section. \ref{sec:level3} where we introduced the details of Quantum Random Access Encoding.
In Sec. \ref{sec:level4} we explain our approach to using Linear Relaxation and in Section. \ref{sec:level7} we present the results of the comparison of QAOA and QRAO on the MKP problem and the performance of QRAO with Linear Relaxation on the Risk-Aware Procurement Optimization problem. Finally in Section.\ref{sec:level8} we provide the concluding remarks and the future outlook of the work.

This research will contribute valuable insights into the practical applicability of quantum and classical hybrid optimization methods for complex optimization challenges in finance and supply chain management.

\section{\label{sec:level2} Problem Settings}

We consider two problem settings. First, the MKP, which is a strongly NP-hard problem with no fully polynomial-time approximation scheme (FPTAS) (unlike the standard Knapsack Problem) \cite{FPTAS}. On this problem, we compare the QAOA and QRAO performance on multiple randomly generated test instances. Next, we consider a real-world supply chain problem akin to an MKP but with an added risk dimension. This is a significantly larger problem, which cannot be solved via QAOA due to the current limit of hardware and memory. The two problems are described as follows: 

\subsection{Multiple Knapsack Problem (MKP)}

The Multiple Knapsack Problem(MKP) \cite{mkp} is a generalization of the standard knapsack problem (KP) from a single knapsack to $m$ knapsacks with (possibly) different capacities. MKP involves allocating a subset of $n$ items to $m$ different knapsacks, aiming to maximize the total profit of the selected items while ensuring that the capacity of each knapsack is not exceeded. This problem finds applications in various fields such as financial management and supply chain, where resources need to be efficiently allocated while considering capacity constraints.

We are given $n$ items that need to be distributed among $m$ knapsacks, each with a distinct capacity $c_i$ for $i = 1, \ldots, m$. Each item $j$ has an associated profit $p_j$ and weight $w_j$. The objective is to select $m$ disjoint subsets of items, ensuring that the items in subset $i$ fit within the capacity $c_i$ of knapsack $i$ while maximizing the total profit of the selected items. Formally, the MKP can be formulated as an Integer Linear Programming (ILP) problem as follows:

\begin{equation}
\begin{aligned}
    \text{Maximize:} \quad & \sum_{i=1}^{m} \sum_{j=1}^{n} p_j x_{ij}  \\
    \text{Subject to:} \quad & \sum_{j=1}^{n} w_j x_{ij} \leq c_i, \quad i = 1, \ldots, m  \\
    & \sum_{i=1}^{m} x_{ij} \le 1 \quad j = 1, \ldots, n, \\
    & x_{ij} \in \{0,1\}, \quad i = 1, \ldots, m, \quad j = 1, \ldots, n 
\end{aligned}
\label{eq:max_constraints}
\end{equation}

where $x_{ij} = 1 $ if item $j$ is assigned to knapsack $i$ , and $x_{ij} = 0$ otherwise. It is usual to assume that the coefficients $p_j, w_j$, and $c_i$ are positive integers, and to avoid trivial cases we demand that:
\begin{equation}
\begin{aligned}
    \max_{j=1,\ldots,n} w_j &\le \max_{i=1,\ldots,m} c_i \\
    \min_{i=1,\ldots,m} c_i &\ge \min_{j=1,\ldots,n} w_j \\
    \sum_{j=1}^n  w_j &\ge \max_{i=1,\ldots,m} c_i
\end{aligned}
\label{eq:demands}
\end{equation}
The first inequality ensures that each item $j$ fits into at least one knapsack as otherwise it may be removed from the problem. If the second inequality is violated, then we may discount the smallest
knapsack, as no items fit into it. The final inequality prevents a trivial solution where all items can fit into the largest knapsack.

\subsection{Risk-Aware Procurement Optimization}

Optimization techniques have found wide application in addressing various challenges within supply chain and logistics management, particularly in mitigating risks and managing uncertainties. For instance, the classic Newsvendor Problem \cite{newsvendor_problem} tackles the dilemma of making purchasing decisions amidst uncertain demand, while disruptions in the supply chain can be managed through portfolio approaches that assess the impact of delays across multiple periods \cite{supply_chain_disruption}. In scenarios where the ramifications of disruptions can be precisely quantified, stochastic optimization techniques are typically employed, solving deterministically across numerous generated scenarios \cite{stochastic_optimization}. However, when data limitations hinder the accurate realization of scenarios for optimization, methods that directly incorporate risk into the optimization process become essential. This can involve integrating risk minimization as a joint objective alongside cost minimization or constraining cost minimization problems with risk considerations \cite{risk_constrained_optimization}.

Supply chain disruptions, whether due to unforeseen events like the COVID-19 pandemic or routine issues such as labor disputes and adverse weather, present significant risks to global companies, leading to delivery delays, missed orders, and financial losses. Addressing these challenges requires a comprehensive approach combining accurate risk quantification with cost-effective decision-making. Following the methodology in \cite{risk}, we establish a supplier risk score metric by analyzing various data sources and identifying key risk factors through factor analysis. Using these risk scores, we develop a risk-constrained optimization model to formulate strategic procurement plans for multinational computer manufacturers. We present a simplified model for a typical procurement setting involving multiple parts and suppliers.

We assign a risk score $r_i \in [0,9]$, for each supplier, $i$, derived from Supplier Risk Analysis \cite{risk}. The per part cost for procuring part $j$ from supplier $i$ is known in advance and denoted as $c_{i, j}$. For simplicity we assume every supplier can produce all parts, in the case where a supplier doesn't produce certain parts, the respective costs will be set to a large value. For each part $j$, there is a demand $d_j$ to be fulfilled by sourcing from multiple suppliers and a risk tolerance level $\psi_j$. The objective is to choose suppliers to fulfill the demands for each part at minimized costs while managing the risk of suppliers within a given tolerance. Eq. \eqref{eq:rbdo_demands} represents the demand constraint and Eq. \eqref{eq:rbdo_risks} ensures the weighted average risks for each part are within the given thresholds.

It is important to note that while the MKP is inherently a maximization problem—where the objective is to maximize the total profit of the selected items—the risk-aware procurement problem, in contrast, is a minimization problem. In the latter, the goal is to minimize the total cost associated with procurement.

\begin{table}[h!]
    \centering
        \caption{Notations used in the model}
        \label{tab:notations}
            \begin{tabular}{|l|p{0.8\linewidth}|}
            \hline
            \textbf{Symbol} & \textbf{Description} \\ \hline
                $i$ & Supplier index from set $\mathcal{I}$ \\ \hline
                $j$ & Part index from set $\mathcal{J}$ \\ \hline
                $y_{i,j}$ & Number of part $j$ obtained from supplier $i$ \\ \hline
                $r_i$ & Risk score for supplier $i$ \\ \hline
                $\psi_j$ & Risk tolerance level for part $j$ \\ \hline
                $c_{i,j}$ & Per part cost of procuring part $j$ from supplier $i$ \\ \hline
                $d_j$ & Demand for part $j$ \\ \hline
            \end{tabular}
\end{table}

Following the notations given in Table~\ref{tab:notations}, the following is the MILP formulation:

\begin{equation}
   \text{Minimize:} \quad \sum_{i \in \mathcal{I}} \sum_{j \in \mathcal{J}} c_{i,j} y_{i,j}
\end{equation}

\begin{align}
   \text{Subject to:} \quad  \sum_{i\in \mathcal{I}} y_{i,j} &\geq d_j && \forall j \in \mathcal{J} \label{eq:rbdo_demands}\\
    \sum_{i\in \mathcal{I}} r_i y_{i,j} &\leq \psi_j d_j && \forall j \in \mathcal{J} \label{eq:rbdo_risks}
\end{align}

\section{\label{sec:level3}Quantum Random Access Codes }

The representation of $n$ qubits as a vector in $\mathbb{C}^{2n}$ may initially suggest a higher information capacity compared to classical $n$ bits. However, it is worth noting that according to the Holevo bound \cite{holevo}, transferring $n$-bit classical information without error requires $n$ qubits. Nevertheless, if some errors are permissible, it becomes possible to encode multiple classical bits into a single qubit using $(n, 1, p)$-QRAC codes \cite{QRAC}. In this context, $(n, m, p)$-QRAC refers to quantum random access codes that encode $n$ classical bits into $m$ qubits.

\begin{mydefinition}{Definition 1}
An $(n,1,p)$-QRA coding is a function that maps $n$-bit strings 
$x \in \{0,1\}^n$ to $1$-qubit states $\rho_x$ satisfying the following condition that for every $i \in \{1,2,...,n\}$ there exists a POVM 

$$ E^{i} = \{E_0^i, E_1^i \}$$

such that 

$$ \text{Tr}(E^i_{x_i}\rho_x) \ge p$$

\end{mydefinition}
The POVM $E^i$ is integral to the decoding process, enabling the extraction of the $i$-th encoded bit $x_i$ from the measured encoded state $\rho_x$ with probability $p$. It's important to note that $(n,1,p)$-QRA codes lack significance when $p \leq \frac{1}{2}$, as $p=\frac{1}{2}$ implies random selection of binary bits. Additionally, $(n,m,p)$-QRA coding, where $m\geq 2$, follows a similar definition. Notably, specific instances include $(2,1,0.85)$- and $(3,1,0.78)$-QRA coding employed in QRAO \cite{QRAO}. However, further extension to $(n,1,p)$ coding with $n\geq 4$ and $p>\frac{1}{2}$ is impractical. This limitation arises from the geometric constraint where a three-dimensional ball cannot be partitioned into sixteen non-empty regions using only four planes \cite{four}.

\begin{myexample}{1} $(3,1,0.78)$-QRA Encoding

Considering the map:
\begin{align}\label{eq:31p}
    & (x_1 ,x_2, x_3) \mapsto \rho_{x_1, x_2, x_3} \notag \\ 
    & :=  \frac{1}{2}\bigg( I + \frac{1}{\sqrt{3}} \Big( (-1)^{x_1} X + (-1)^{x_2} Y + (-1)^{x_3}Z \Big) \bigg)
\end{align}

For every pair of $(x_1, x_2, x_3), \rho_{x_1, x_2, x_3}$ is a pure state and can be written in the form 
$\rho_{x_1, x_2, x_3} = |\psi(x_1, x_2, x_3)\rangle \langle \psi(x_1, x_2, x_3)|$, where:

\begin{equation}
\begin{aligned}
    |\psi(0,0,0)\rangle & = \cos \tilde{\theta}|0\rangle + e^{\frac{\pi \iota}{4}} \sin\tilde{\theta} |1\rangle \\
    |\psi(0,0,1)\rangle & = \sin \tilde{\theta}|0\rangle + e^{\frac{\pi \iota}{4}} \cos\tilde{\theta} |1\rangle \\
    |\psi(0,1,0)\rangle & = \cos \tilde{\theta}|0\rangle + e^{\frac{-\pi \iota}{4}} \sin\tilde{\theta} |1\rangle \\
    |\psi(0,1,1)\rangle & = \sin \tilde{\theta}|0\rangle + e^{\frac{-\pi \iota}{4}} \cos\tilde{\theta} |1\rangle \\
    |\psi(1,0,0)\rangle & = \cos \tilde{\theta}|0\rangle + e^{\frac{3\pi \iota}{4}} \sin\tilde{\theta} |1\rangle \\
    |\psi(1,0,1)\rangle & = \sin \tilde{\theta}|0\rangle + e^{\frac{3\pi \iota}{4}} \cos\tilde{\theta} |1\rangle \\
    |\psi(1,1,0)\rangle & = \cos \tilde{\theta}|0\rangle + e^{\frac{-3\pi \iota}{4}} \sin\tilde{\theta} |1\rangle \\
    |\psi(1,1,1)\rangle & = \sin \tilde{\theta}|0\rangle + e^{\frac{-3\pi \iota}{4}} \cos\tilde{\theta} |1\rangle 
\end{aligned}    
\label{eq:complex}
\end{equation}

where $\tilde{\theta}$ satisfies the condition $(\cos(\tilde{\theta})^2 = \frac{1}{2} + \frac{1}{2\sqrt{3}} > 0.78$. Then this map is a $(3,1,0.78)$-QRA codings with the POVMs 
\begin{align}
    E^{1} = \{|+\rangle \langle +| , |-\rangle \langle -|\}, & \   E^{2} = \{|+\iota\rangle \langle +\iota| , |-\iota\rangle \langle -\iota|\}\notag 
    \\
    E^{3} = & \{|0\rangle \langle 0|  , |1\rangle \langle 1|\} \label{eq:povm}
\end{align}

\end{myexample}

The measurements described above are conducted in the $X$, $Y$, and $Z$ bases. Each measurement is specifically performed to decode the corresponding classical bit encoded within the quantum state.

Utilizing QRAO \cite{QRAO}, multiple classical bits are compactly encoded into a reduced number of qubits using Quantum Random Access Codes (QRACs), as explained earlier. For instance, employing a $(3, 1)$-QRAC, three classical binary variables $x_1$, $x_2$, and $x_3$ are mapped to a single qubit through the application of the Pauli $X$, $Y$, and $Z$ operators, respectively. In comparison to methods like QAOA or VQE, QRAO boasts a constant-factor space complexity advantage. Consequently, our focus in this study lies on leveraging QRAO with a (3, 1)-QRAC approach.

The objective is to simplify the optimization problem by directing it towards the exploration of the maximum eigenstate of the relaxed Hamiltonian $H_{\text{relax}}$. To achieve this, we first map classical binary variables into qubits through the construction of a relaxed Hamiltonian. This process involves performing a graph coloring of the instance graph $G$, (made from the objective problem) using the Large Degree First (LDF) method \cite{ldf}, with time complexity of $O(|V(G)| \log |V(G)| + \text{deg}(G)|V(G)|)$. Upon completion of the LDF algorithm, the vertices are partitioned into the set $V_c$ associated with the color $c \in C$. Here, the color of the $i$-th vertex $v_i$ is denoted as color($i$). This partitioning satisfies the following condition:

\begin{equation*}
    e_{i,j} \in E(G) \Rightarrow \text{color}(i) \ne \text{color}(j)
\end{equation*}

Next, we allocate $\ceil[\Big]{\frac{|V_C|}{3}}$ qubits for each color $c \in C$, enabling up to three vertices to be assigned to a single qubit. These vertices are ordered, and the Pauli operators $X$, $Y$, and $Z$ are assigned accordingly. Subsequently, we employ variational methods like VQE to explore the maximum eigenstate of $H_{\text{relax}}$. Unlike the diagonal structure of the original Hamiltonian, $H_{\text{relax}}$ comprises non-classical states, characterized by superposition and entanglement, as its maximal eigenstates. Consequently, the eigenstate obtained for $H_{\text{relax}}$ cannot be directly linked to the classical solution due to its quantum nature. Instead, it represents a quantum state corresponding to the relaxed solution of the optimization problem, where the constraint that the solution must be a binary vector is lifted. To recover the classical solution, we employ quantum state rounding algorithms, as proposed in \cite{QRAO}.

The first rounding algorithm, Pauli rounding, deciphers the encoded three classical bits in each qubit using the POVM outlined in Eq. \ref{eq:povm}. Essentially, this procedure involves measuring the $j$-th qubit with sufficient repetitions, determining the majority measurement outcome, and assigning it as the rounded value of the corresponding classical bit. However, Pauli rounding may encounter limitations when the relaxed state exhibits significant entanglement, preventing its representation in the form of $\rho_1 \otimes \rho_2 \otimes \ldots \otimes \rho_n$. In such cases, the algorithm's effectiveness may be compromised due to the oversight of correlations among the qubits.

In contrast to Pauli rounding, the second rounding algorithm, known as Magic Rounding, mitigates the aforementioned issue. This method aims to decode three classical variables simultaneously from a single qubit. For a comprehensive understanding of Magic Rounding, further details can be found in \cite{QRAO} and \cite{Teramoto22023}.

\section{\label{sec:level4}Proposed Approach}

An integer linear program (ILP) consists of a linear program requiring variables to be integers. ILPs serve as expressive tools for formulating combinatorial optimization problems. Nonetheless, solving ILPs optimally is NP-hard.

One standard method to find an approximate solution for a combinatorial optimization problem is via Linear Relaxation: 

\begin{itemize}
    \item Formulate the optimization problem as an ILP.
    
    \item Derive a linear program (LP) from the ILP by relaxing the integrality constraints on variables. This resulting LP termed a relaxation of the original problem, retains the same objective function but operates over a broader solution set, leading to $\textit{opt}(LP) \leq \textit{opt}(ILP)$ for minimization problem.
    
    \item Solve the LP optimally using an efficient linear programming algorithm and rounding variables to integers by some rounding techniques. 
\end{itemize}

In the current NISQ era, quantum algorithms still suffer from scaling to large-size problems. In our approach, we apply linear relaxation as a method to reduce the problem size. Given the LP relaxed solution, the binary decision variables that are solved to extreme values, i.e. very close to 0 or 1, will be rounded accordingly and fixed, leaving a reduced-size problem for QRAO to tackle.

\section{\label{sec:level7}Results}

We performed a series of preliminary investigations to identify the optimal ansatz and associated parameters for employing $(3,1,p)-$QRAO. These investigations were carried out by solving a simplified variant of our target problem. All experiments utilized Qiskit's AerSimulator \cite{Qiskit}. Specifically, we conducted $20$ experiments on small-scale Multi Knapsack Problems (MKP), each consisting of three bins and three items. Each bin possessed a distinct capacity $(c_i)$, while each item had unique values $(p_j)$ and weights $(w_j)$, resulting in approximately $10$ binary variables and six constraints per problem instance. The weights were randomly assigned within the range $ w_j \in [1,3]$, the item values within $ p_j \in [1,10]$, and the bin capacities within $c_i \in [1,3]$. The primary metric of interest was the frequency of solutions achieving an optimality gap of less than $0.05\%$.

We employed a variational ansatz incorporating the brickwork architecture, as illustrated in Fig. \ref{fig:quantum_circuit}. This ansatz features alternating layers of single-qubit rotations and $R_{XX}$ gates. Each layer of single-qubit gate contains rotations around a single direction ($X$, or $Y$, or $Z$), one at a time, sequentially followed by a layer of two-qubit rotation gates, each controlled by its variational parameter. This design enhances flexibility and facilitates detailed exploration of the parameter space through its structured sequence of operations.

\begin{figure}[ht]
    \centering
    \resizebox{0.5\textwidth}{!}{
        \begin{quantikz}
            \lstick{\ket{0}}&\gate{R_X(\theta_1)}& \gate[2]{R_{XX}(\theta_5)}&\gate{R_Y(\theta_7)}&& \gate{R_Z(\theta_{12})}&\gate[2]{R_{XX}(\theta_{16})}& \\
            \lstick{\ket{0}}&\gate{R_X(\theta_2)}&&\gate{R_Y(\theta_8)}&\gate[2]{R_{XX}(\theta_{11})}&\gate{R_Z(\theta_{13})}&& \\
            \lstick{\ket{0}}&\gate{R_X(\theta_3)}&\gate[2]{R_{XX}(\theta_6)}&\gate{R_{Y}(\theta_9)}&& \gate{R_Z(\theta_{14})}& \gate[2]{R_{XX}(\theta_{17})} & \\
            \lstick{\ket{0}}&\gate{R_X(\theta_4)}&&\gate{R_Y(\theta_{10})}&&\gate{R_Z(\theta_{15})}&& 
        \end{quantikz}
    }
    \caption{Variational ansatz as a brickwork architecture, with layers of single-qubit rotations around a single direction ($X$, or $Y$ or $Z$), one at a time, and a layer of two-qubit rotation gate ($R_{XX}$). This quantum circuit shows three complete layers of BrickWork ansatz.}
    \label{fig:quantum_circuit}
\end{figure}

Our results in Tab. \ref{tab:performance} consistently indicate a preference for the \textit{BrickWork} and \textit{EfficientSU2} ansatz (full entanglement) over alternatives such as \textit{PauliTwoDesign} and \textit{RealAmplitudes}. This preference is informed by the intrinsic properties outlined in Eq. \ref{eq:complex}, which reveals that the $(3,1,p)-$QRAC framework utilizes complex amplitudes. In contrast, the \textit{RealAmplitudes} ansatz relies solely on real-valued parameters.

The \textit{BrickWork} and \textit{EfficientSU2} ansatz are optimized for hardware efficiency within \textit{SU(2) 2-local} circuits. These circuits consist of layers of single-qubit operations interconnected by \textit{SU(2)} and \textit{CX} entanglements, where \textit{SU(2)} refers to the special unitary group of degree 2, represented by $2 \times 2$ unitary matrices with a determinant of 1. Among these, \textit{BrickWork} demonstrates a slight performance advantage due to its additional variational parameters while retaining all features of \textit{EfficientSU2}.

\begin{table}[h!]
    \centering
    \begin{tabular}{|c|r|r|r|r|r|r|r|r|r|r|r|r|}
        \hline
        \textbf{Layers} & \textbf{B.W} & \multicolumn{3}{c|}{\textbf{EfficientSU2}} & \multicolumn{3}{c|}{\textbf{Pauli 2 Design}} & \multicolumn{3}{c|}{\textbf{Real Amps.}} \\ \hline
         & Default & Circ & Full & Lin & Circ. & Full & Lin & Circ. & Full & Lin. \\ \hline
        0 & 12 & 9 & 11 & 8 & 8 & 12 & 8 & 0 & 2 & 6 \\ \hline
        1 & 15 & 11 & 16 & 8 & 10 & 15 & 8 & 1 & 3 & 0 \\ \hline
        2 & 18 & 11 & 15 & 12 & 12 & 14 & 10 & 1 & 2 & 0 \\ \hline
        3 & 20 & 15 & 18 & 16 & 13 & 16 & 15 & 1 & 4 & 2 \\ \hline
        4 & 20 & 16 & 18 & 18 & 15 & 16 & 15 & 2 & 4 & 2 \\ \hline
    \end{tabular}
    \caption{Number of solved instances for various ansatz, employing different types of entanglement, with increasing layers (\textbf{B.W.} = BrickWork ansatz).}
    \label{tab:performance}
\end{table}

\subsection{QAOA vs QRAO on Multiple Knapsack Problem}

Subsequently, we conducted a comparative analysis between QAOA and $(3,1,p)-$QRAO, evaluating their performances on $100$ randomly generated, non-trivial (guaranteed by Eq. (\ref{eq:demands})) instances of the MKP.
The instances varied in complexity, with the number of bins ranging from $2$ to $5$ and the number of items from $2$ to $10$. The weights $w_j$ of items and capacities $c_i$ of the bins, as well as the values of the items $p_j$, were assigned randomly within the intervals $[1,3]$,$[1,3]$ and $[1,10]$, respectively, with six constraints.
The problem sizes comprised a maximum of $20$ binary variables.

\begin{figure}[H]
    \centering
    \includegraphics[width=\linewidth]{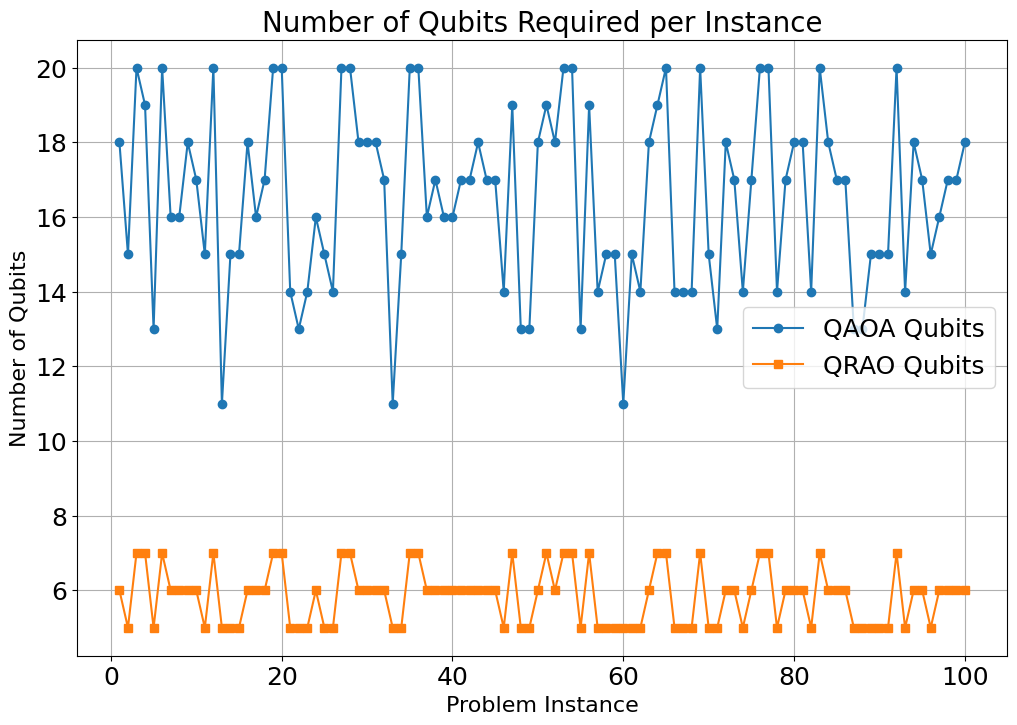}
    \caption{Comparison on number of qubits required for each of the $100$ problem instances in both QAOA and QRAO method. }
    \label{fig:qubit_comparison}
\end{figure}

In QAOA, each variable necessitated one qubit, while QRAO utilized fewer qubits due to the compression of classical information in qubit amplitudes, as seen in Fig. \ref{fig:qubit_comparison}.

In QAOA, we employed the default \textit{QAOA Ansatz} with $5$ layers. Conversely, for the $(3,1,p)$-QRAO algorithm, we employed the BrickWork Ansatz, configured with $8$ complete layers. Each complete layer of the BrickWork Ansatz comprises a single layer of single-qubit rotations and a single layer of $R_{XX}$ gates. In both cases, we used COBYLA as the classical optimizer. Both the Pauli rounding and Magic rounding schemes were employed for QRAO, where Magic Rounding performed better. 
QAOA, when implemented with $20$ qubits, was generally constrained in terms of circuit depth, limiting its performance and scalability. This restriction hindered the exploration of more complex optimization landscapes and the achievement of higher solution quality. On the other hand, QRAO, utilizing an average of $7$ qubits for the same problem, demonstrated the capability to achieve significantly deeper circuits. This increased depth allowed for more extensive computational processes, potentially leading to better optimization results and greater efficiency in solving complex problems. The difference in qubit requirements highlights the potential advantages of QRAO in quantum optimization tasks, particularly in environments with limited qubit resources, as also shown in Tab. \ref{tab:mkp-results}.

The optimal objective values, obtained by CPLEX, were used to compare with the solutions obtained from QAOA and QRAO. The average optimality gap is also calculated, which is used to evaluate the performance of an optimization algorithm. It represents the average difference between the objective value obtained by the algorithm and the optimal objective value, relative to the optimal value, across a set of problem instances.
\begin{table}[htbp]
    \centering
    \begin{tabular}{|p{1.5cm}|p{1.5cm}|>{\raggedleft\arraybackslash}p{1.5cm}|>{\raggedleft\arraybackslash}p{1.5cm}|>{\raggedleft\arraybackslash}p{2cm}|}
        \hline
        Method & Rounding & Feasible & Optimal & Optimality Gap \\
        \hline
        QAOA & - & $98\%$ & $63\%$ & 0.092\\
        \hline
        QRAO & Pauli & $81\%$ & $4\%$ & 0.606\\
        \hline
        QRAO & Magic & $98\%$ & $70\%$ & 0.065\\
        \hline
    \end{tabular}
    \caption{MKP performance comparisons of QAOA vs QRAO results}
    \label{tab:mkp-results}
\end{table}

\subsection{QRAO with Linear Relaxation on Risk-Aware Procurement Optimization Problem}
We generated multiple random instances of the Risk-Aware Procurement Optimization problem \cite{risk}, each consisting of $\geq 100$ binary variables. The scale of these problems exceeds the current hardware limitations of QAOA, making them difficult to solve directly. Utilizing IBM's Qiskit simulator, we were limited to the \textit{matrix product state} simulator, which supports a maximum of 63 qubits for running our problems.

As problem size increases, QRAO's efficiency is similarly impacted. The reduction in qubits necessitates greater circuit depth to maintain expressivity. However, running circuits with more than 50 qubits to achieve this depth is impractical due to memory constraints.

We employed the Linear Relaxation technique as a strategic response to address these challenges. We tested this approach on 10 problem instances containing at least $100$ binary variables. Initially, we experimented by fixing a random subset of variables. Specifically, variables whose values were within a range of \(\delta\) from 0 were fixed to 0, and those within a range of \(\delta\) from 1 were fixed to 1. Here, \(\delta\) represents a threshold value that determines how close a variable's continuous value must be to 0 or 1 before it is fixed to those binary values. For instance, if \(\delta = 0.1\), variables with values in the interval \([0, 0.1)\) were fixed to 0, while variables with values in the interval \((0.9, 1]\) were fixed to 1. This method was compared to the scenario where 90\% of the variables were fixed directly. The use of \(\delta\) helps to manage the complexities introduced by increasing problem dimensions by focusing on variables that are already close to their boundary values, thus simplifying the problem space.

By fixing a significant portion of the variables, we effectively reduce the dimensionality of the problem, which in turn alleviates the memory and computational burdens associated with large-scale instances. This approach not only makes it feasible to handle problems that would otherwise exceed current hardware limitations but also provides insights into the behavior of the optimization process under different levels of variable fixation. This strategy serves as a viable pathway forward, offering a practical solution for navigating the constraints of available computational resources while still addressing the core challenges of the Risk-Aware Procurement Optimization problem.

\begin{table}[htbp]
    \centering
    \begin{tabular}{|p{1.5cm}|p{1.5cm}|>{\raggedleft\arraybackslash}p{1.5cm}|>{\raggedleft\arraybackslash}p{1.5cm}|>{\raggedleft\arraybackslash}p{1.5cm}|}
        \hline
        Method & Fixed \%age & Feasible & Optimal & Optimality Gap \\
        \hline
        QRAO & Random & 100\% & 80\% & 0.053\\
        \hline
        QRAO & $90\%$ & 70\% & 50\% & 0.088\\
        \hline
        QRAO & $85\%$ & 70\% & 50\% & 0.072\\
        \hline
    \end{tabular}
    \caption{QRAOs' (with Linear Relaxation) performance comparison for different $\%$age of fixed variables.}
    \label{tab:qrao-lr-results}
\end{table}

\section{\label{sec:level8}Conclusion \& Future Outlook}
This paper introduces a hybrid methodology for addressing constraint optimization problems in business contexts, focusing on MKP and MKP with an additional risk dimension. Our analysis shows that QRAO, using fewer qubits, achieves performance comparable to QAOA and enables solving problems with over 100 binary variables. This scalability is significant as QAOA faces challenges from high qubit and memory demands at such scales.

Moreover, our study demonstrates that by integrating the classical technique of Linear Relaxation, we can achieve a further reduction in qubit requirement. This integration not only minimizes the resources needed but also enhances the quality of the obtained results. By leveraging Linear Relaxation alongside quantum approaches, our method offers a comprehensive solution framework that optimizes both resource utilization and solution efficacy. This combination underscores the versatility and effectiveness of our hybrid approach in tackling complex optimization challenges across diverse business domains.

Expanding on this study, we aim to enhance our approach by integrating stochastic elements into the problem formulation. This entails addressing uncertainty surrounding both demand and supply variables \cite{news}. Furthermore, we will investigate the potential of employing various classical methods to streamline the problem-solving process. 

These adaptations are anticipated to broaden the scope of our methodology and enhance its effectiveness across diverse problem landscapes.

\section{Acknowledgement}

This research is supported by the National Research Foundation, Singapore under its Quantum Engineering Programme 2.0 (NRF2021-QEP2-02-P01).

\end{document}